\documentstyle{amsppt} \TagsAsMath \magnification=\magstep1 \hsize=14truecm  \vsize=21truecm  \def\P{\Bbb P} \def\E{\Bbb E} \def\RE{\Bbb R} \def\L{L^2(\RE^3)} \def\l{\lim_{t\uparrow \infty }\,} \def\p{\par\noindent} \def\v{\vskip 8pt\p}  \def\C{{\Bbb C}} \def\F{{\Cal F}} \def\qed{\hfill\vbox{\hrule\hbox{\vrule\vbox to 7 pt {\vfill\hbox to          7 pt {\hfill\hfill}\vfill}\vrule}\hrule}\par} \def\uno{1} 
 \centerline{{\bf SCATTERING INTO CONES}} \centerline{{\bf AND FLUX ACROSS SURFACES}} \centerline{{\bf IN QUANTUM MECHANICS:}}  \centerline{{\bf A PATHWISE PROBABILISTIC APPROACH}} \vskip 60pt \centerline{Andrea Posilicano } \centerline{Dipartimento di Scienze, Universit\`a dell'Insubria} \centerline {Via Valleggio 11, I-22100 Como, Italy} \centerline{E-mail: {\tt andreap\@uninsubria.it}} \vskip 20pt \centerline{Stefania Ugolini} \centerline{Dipartimento di Matematica, Universit\`a di Milano} \centerline{Via Saldini 50, I-20133 Milano, Italy} \centerline{E-mail: {\tt ugolini\@mat.unimi.it}} \vskip 120pt \noindent {{\bf Abstract.}} { We show how the scattering-into-cones and flux-across-surfaces theorems in  Quantum Mechanics have very intuitive pathwise probabilistic versions based  on some results by Carlen about large time behaviour of paths of  Nelson's diffusions.  The quantum mechanical results can be then recovered by  taking expectations in our pathwise statements.}  
\vfill\eject \p{\bf 1. Introduction.} \v The problem of finding the basic mathematical relationships between theoretical previsions and experimental observable quantities has been, for a long time, an open problem in quantum theory of scattering.\par In this direction there exists two relevant theorems. The first one is due to Dollard in 1969 (see [D]) and states that the probability of asympotically observing the particle in some cone $C\subset\RE^3$ with  vertex in the scattering center is equal to the probability of finding its asymptotic momentum exactly in the same cone, i.e. $$ \l \int_{C}dx\,|\psi_t(x)|^2=\l \int_{C\cap B^c_R}dx\,|\psi_t(x)|^2= \int_C dk\,|\widehat\psi_{\text{\rm out}}(k)|^2\ , \eqno(1.1) $$ where $B_R^c$ is the complement of $B_R$, the ball of radius $R$,  $\widehat{\ }$ denotes the Fourier transform and $\psi_{\text{\rm out}}:= \Omega_+^*\psi_0$ is the outgoing state, $\Omega_+$ being the wave operator. It is well known that the differential cross section for the time-independent scattering theory can be derived from the right hand side of (1.1). Nevertheless the importance of (1.1) is primarily conceptual since the probability of observation which it refers to is a time-asymptotic one.\par  Instead, in the usual experimental situation, the detector being sufficiently far away from the scattering centre, one actually measures the probability that the particle crosses the active surface of the detector $C\cap S_R$ ($S_R$ denoting the sphere of radius $R$) at some random time. The theorem which takes care of this experimental setting is the so-called flux-across-surfaces theorem. It was conjectured in 1975 by Combes, Newton and Shtokhamer (see [CNS]) under the form of the following relation: $$ \lim_{R\uparrow \infty}\,\int_{t_0}^{+\infty}dt\, \int_{{C\cap S_R}}d\sigma(x)\,J^{\psi_t}(x)\cdot n(x) =\int_Cdk\, |\widehat\psi_{\text{\rm out}}(k)|^2\ , \eqno (1.2) $$   where $J^{\psi_t}:=\text{\rm Im}\,\psi_t^*\nabla\psi_t$ is the quantum probability current density, $n$ denotes the outward unit normal  vector along $C\cap S_R$  and $\sigma$ is the surface measure. \par  No rigorous proof of this conjecture was known  until 1996 when Daumer, D\"urr, Goldstein and Zangh\`\i\ (see [DDGZ1]) proved the flux-across-surfaces theorem in the free case. Successively the result has been extended to the interacting case by Amrein and Zuleta (see [AZ])  and by Teufel, D\"urr and M\"unch-Berndl (see [TDM-B]) for short range potentials, and by Amrein and Pearson (see [AP]) for long range potentials. The case with zero-energy resonances or  eigenvalues  has been treated by Dell'Antonio and Panati in [DPa]  and the case with a delta interaction by Panati and Teta in [PT].  \par In view of our approach, the most interesting proof is the one given  in [AP]. From such a paper one can extract the following clarifying  scheme:\par Integrating with respect to time the equation of continuity for quantum probability density   $$ \frac{\partial}{\partial t}\,|\psi_t|^2+\nabla  J^{\psi_t}=0\eqno(1.3) $$ and inserting the result into the relation (1.1) given by  Dollard's theorem one obtains $$ \int_C dk\,|\widehat\psi_{\text{\rm out}}(k)|^2= \int_{C\cap B_R^c}dx\,|\psi_{t_0}(x)|^2\ -\  \l\int^t_{t_0}ds\int_{C\cap B_R^c}dx\,\nabla J^{\psi_{s}}(x)\,. $$ Then by taking the limit $R\uparrow\infty$ and by Gauss-Green divergence theorem one has $$ \int_C dk\,|\widehat\psi_{\text{\rm out}}(k)|^2= \lim_{R\uparrow\infty}\,\l\int_{t_0}^tds\int_{(C\cap S_R)\cup(\partial C\cap B^c_R)}d\sigma(x)\,n(x)\cdot J^{\psi_s}(x)\,, $$ and so the flux-across-surfaces theorem is a consequence of the scattering-into-cones theorem plus the condition $$ \lim_{R\uparrow\infty}\,\l\int_{t_0}^tds\int_{\partial C\cap B^c_R}d\sigma(x)\,n(x) \cdot J^{\psi_s}(x)=0\,\eqno (1.4) $$ i.e. the flux across the lateral boundary of the cone asymptotically  vanishes. \par In this paper we give a pathwise formulation of scattering-into-cones  and flux-across-surfaces theorems following in some way the pathwise  analogue of the above analytic argument. This has the advantage of giving a pictorial view of the scattering behaviour.  In doing that we exploit the relevant  results, obtained by Carlen in 1985, about potential scattering in Stochastic Mechanics (see [C4]). \par It is known that Stochastic Mechanics, introduced by Nelson in 1966  (see [N1-3]), allows a pathwise approach to Quantum Mechanics by providing a suitable class of diffusion processes. Indeed to a solution  $\psi_t$ of the Schr\"odinger equation there is associated a well defined (see Theorem 1) diffusion process $X_t$ solution of the stochastic differential equation  $$dX_t=b(t,X_t)\,dt+dB_t$$  where $B_t$ is a Brownian motion and the drift vector  field $b_t(x)\equiv b(t,x)$ is given by $$ b_t=|\psi_t|^{-2}(\nabla|\psi_t|^2+J^{\psi_t})\ . $$ Moreover the probability density of the process $X_t$ is given by $|\psi_t|^2$ and it satisfies the continuity (or Fokker-Planck) equation (1.3).  In connection with the problem of potential scattering, Carlen studied the time evolution of the process $\frac{1}{t}\,X_t$ proving that (see Theorem 3): \smallskip \item{1.} the scattering diffusions (i.e. the ones associated to the scattering states of the corresponding Schr\"odinger equation) are such that the limit $$ \l\frac{1}{t}\, X_t=p_+  $$ exists almost surely; \smallskip \item{2.} the random variable $p_+$ is square integrable and has the same distribution as does the quantum mechanical final momentum. \smallskip\p These facts imply  that almost surely the diffusion paths are definitively inside or outside  the cone $C$, a pathwise analogue of (1.4). Then the following pathwise version of Dollard's theorem immediately follows:  $$\l \chi_{C}( X_t)= \l \chi_{C\cap B_R^c}( X_t)=\chi_C( p_+)\,, $$ $\chi_D$ denoting the characteristic function of the set $D$;  the usual quantum mechanical version is then obtained  by taking  expectations (see Theorems 4 and 5). \par As regards the flux-across-surfaces theorem the situation is almost equally simple. If $N_{C\cap S_R}$ were finite, where  $$ N_{C\cap S_R}:= N^+_{C\cap S_R}-N^-_{C\cap S_R}\ ,$$ $N^+_{C\cap S_R}(\gamma)$ (resp. $N^-_{C\cap S_R}(\gamma) $) denoting the number of outward (resp. inward) crossing by the path $ t\mapsto \gamma(t)$ of $C\cap S_R$, then, again by 1 and 2 above, one would obtain the following pathwise version of the flux-across-surfaces theorem: $$ \lim_{R\uparrow\infty}\,N_{C\cap S_R}=\chi_C(p_+)\,.\eqno(1.5) $$ Let us remark here that the relevance of $N_{C\cap S_R}$ for the flux-across-surfaces theorem was already pointed out  (in the framework of Bohmian Mechanics) in [DDGZ2]. The problem here is that  almost surely the diffusion $X_t$ intersects ${C\cap S_R}$ on a set of times that has no isolated point and is uncountable. Therefore the  definition of $N_{C\cap S_R}$ given above makes no sense in general. However, by a suitable redefinition of $N_{C\cap S_R}$ as the total mass of an almost surely  compactly supported random distribution (see section 3 for the details), 1.5 can be made rigorous (see Theorem 6). After showing (see Theorem 7) how to explicitly compute, by using the continuity equation (1.3),  the expectation of $N_{C\cap S_R}$ in terms of the quantum probability current density $J^{\psi_t}$,  the flux-across-surfaces  theorem then follows by taking expectation in (1.5) (see Theorem 9).\par In our opinion these results show how the probabilistic approach we use is very fruitful and extremely intuitive from the physical point of view.  \par  As regards the analytical hypotheses we impose, our proofs of the  pathwise results need, beside the existence of the asymptotic velocity (see hypotheses h.3, h.4 in definition 2), the following  condition on the quantum evolution: $$ \int_{t_0}^{+\infty}\frac{dt}{t}\,\left\|\left(P-\frac{Q}{t}\right)\psi_t\right\|_{L^2} <+\infty\ ,\qquad t_0>0\ .\eqno(1.6) $$ where $P\psi(x):=-i\nabla\psi(x)$ and $Q\psi(x)=x\phi(x)$ denote the usual momentum and position operators of Quantum Mechanics in Schr\"odinger representation. Let us remark that the original results by Carlen were obtained by requiring the existence and completeness of wave operators, which is a hypothesis stronger than our h.3 and h.4. It is not clear to us if our weaker hypotheses together with (1.6) in any case imply existence and completeness of wave operators. Therefore it could be interesting to find examples (if any) of cases in which the pathwise scattering-into-cones and flux-across-surfaces theorems hold true  notwithstanding there are no wave operators.  \par In order to obtain then the quantum mechanical results by taking expectations, (1.6) is  still sufficient to get Dollard's theorem, whereas the  flux-across-surfaces theorem requires that the property of paths of being definitively always inside or outside the cone $C$ holds not only pathwise but in the mean, i.e, as we already know, (1.4) must be true. This condition  is a consequence of  $$ \int_{t_0}^{+\infty}dt\,\left\|\theta(Q)\psi_t\right\|_{H^1}\, \left\|\left(P-\frac{Q}{t}\right)\psi_t\right\|_{H^1}<+\infty\,.\eqno(1.7) $$ where $\theta\in C^2_b(\RE^3;\C)$, $\theta=1$ on a neighbourhood of $\partial C\cap B_R^c$ for some $R>0$ and $H^s(\RE^3)$ denotes the Sobolev space of tempered distributions with a Fourier transform which is integrable with respect to the measure with density $(1+|x|^2)^s$.  \par Conditions (1.6) and (1.7) both follows from  propagation estimates on $\psi_t$. This is a well-known topic in mathematical physics  and a large literature exists on them. Thus by using known results on time-decay of the solutions of the Schr\"odinger equation it is possible to deduce (1.6) and (1.7) from explicit conditions imposed on the initial state $\psi_0$ and on the potential $V$, which are the natural prescriptions for a physicist. In particular, beside some technical condition on the initial state $\psi_0$, (1.6) holds true with potential functions decaying at infinity like $\|x\|^{-\epsilon}$, $\epsilon>0$, whereas (1.7) requires potentials decaying faster that $\|x\|^{-2/3}$ (see section 5  for more details). These  conditions on $\psi_0$ and $V$ also lead to existence and  completeness of (modified) wave operators.  \par Our probabilistic proof remains unchanged in the case of the presence either of a time-dependent potential or of a magnetic field, the only difference, if $A$ denotes the magnetic potential,  being the replacement of $P$ by $P-A$ and of $J^{\psi_t}$ by $J^{\psi_t}-|\psi_t|^2A$. We plan to work out the details in a future work. \par Finally let us remark that all our results hold true every time we can find a stochastic process $X_t$ having $|\psi_t|^2$ as its density and for which Theorem 3 can be proven. By Theorem 1 we realized such a process as a Nelson Diffusion, but this is not the only possible choice. Another one is given by Bohmian Mechanics  (see [DGZ] for a thorough introduction to the subject), where one considers the stochastic process $\tilde X_t$, solution of the ordinary differential equation $\displaystyle \frac{d}{dt}\,\tilde X_t=|\psi_t(\tilde X_t)|^{-2}J^{\psi_t}(\tilde X_t)$ with a random initial condition with density $|\psi_{0}|^2$. Also in this case, under the same  hypotheses plus the techinical condition $\psi_{t_0}\in C^\infty(\RE^3)$ (the Bohmian analogue of Theorem 1, see [BDGPZ], needs more regularity), Theorem 3 holds true, the proof being essentially the same, and so all our results can be stated in a Bohmian context. We decided to work with Nelson's stochastic mechanics since it does not necessarily need the Schr\"odinger equation in  its formulation. Indeed it can be derived either from a stochastic analogue of Newton's law (see [N1], [N2]) or from a stochastic  variational principle (see [GM], [N3]).   \vskip 20pt\p {\bf 2. Potential Scattering in Stochastic Mechanics} \v At first let us recall that, by Nelson's Stochastic Mechanics  (see [N1-3]), it is possible to associate to a solution  $\psi_t$ of  the Schr\"odinger equation a diffusion process which has $|\psi_t|^2$ as  its density. More precisely one has the following (in the reference [C1] $V$ is a Rellich-class potential but the results obtained there can be extended to the more general potentials used here by  proceeding as in [DP, Theorem 2.1])\v {\bf Theorem 1.} ([C1]) {\it Let $V=V_1+V_2$, with $V_1$ bounded from below and $V_2$ $(-\Delta)$-form-bounded with  relative bound smaller than one. Let $H=-\frac{1}{2}\,\Delta+ V$ be defined  as a sum of quadratic forms, and let  $\psi_0$ be a normalized state in its form domain ${\Cal Q}(H)=H^1(\RE^3)\cap {\Cal Q}(V_1)$. If $\psi_t:=e^{-itH}\psi_0$, define then  $$ b(t,x):=b_t(x)\,,\qquad b_t:=|\psi_t|^{-2}(\nabla|\psi_t|^2+J^{\psi_t})\ . $$  Consider the measurable space $(\Omega,\F)$,  with $\Omega=C([t_0,+\infty);\RE^d)$, $t_0\ge 0$, $\F$ the Borel $\sigma$-algebra, and let  $(\Omega,\F,\F_t,X_t)$ be the evaluation stochastic process  $X_t(\gamma):=\gamma(t)$, with $\F_t=\sigma(X_s,\,t_0\le s\le t)$ the natural  filtration. Then there exists a unique Borel probability measure $\P$ on  $(\Omega,\F)$ such that:\p \smallskip\p --$\ $ $(\Omega,\F,\F_t,X_t,\P)$ is a Markov process; \smallskip\p --$\ $ the image of $\P$ under $X_t$ has density $|\psi_t|^2$; \smallskip\p --$\ $ $\displaystyle B_t:=X_t-X_{t_0}-\int_{t_0}^tds\, b(s,X_s)$ is a  $(\P,\F_t)$-Brownian motion, i.e. $\P$ is a weak solution of the  stochastic differential equation $dX_t=b(t,X_t)\,dt+dB_t$ with initial density  $|\psi_{t_0}|^2$. } \v From now on we will assume $t_0>0$ and $d=3$. \v {\bf Definition 2.} With the same notation and hypotheses  as in Theorem 1, let us call the couple  $(\psi_0,V)$ {\it weakly admissible} if \smallskip\p h.1) $\psi_0$ is in $\Cal H_c$, the spectral subspace corresponding to the continuous spectrum of $H$; \smallskip\p h.2) $$\int_{t_0}^{+\infty} \frac{dt}{t}\, \left\|\left(P-\frac{Q}{t}\right)\psi_t\right\|_{L^2} <+\infty\ . $$ \smallskip\p h.3) the asymptotic velocity exists in the following sense: $$ \forall\,g\in C^\infty_c(\RE^3)\,,\qquad \text{\rm w-}\lim_{t\uparrow\infty}\,  \Pi_c\,e^{itH}\,g\left(\frac{Q}{t}\right)\, e^{-itH}\Pi_c =\Pi_c\,g(P_+)\,\Pi_c\,, $$ where $\text{\rm w-}\lim$ means the limit in the weak operator  norm topology, $\Pi_c$ denotes  the projection onto $\Cal H_c$ and $P_+$ is a vector of commuting  self-adjoint operators; \v A weakly admissible couple $(\psi_0,V)$ is then called {\it admissible} if moreover \smallskip\p h.4) $\psi_0$ is in the spectral subspace corresponding to the absolutely continuous spectrum of $P_+$. \v {\bf Remark.} If the (modified) wave operators  $\Omega_\pm$ exist and are complete then $$ P_+\Pi_c=\Omega_+P\Omega_+^*\Pi_c $$ and, given $\psi_0\in{\Cal H}_c$, $$ \left\langle\psi_0,\chi_A(P_+)\psi_0\right\rangle=\langle\psi_0,\Omega_+\chi_A(P)\Omega_+^*\psi_0\rangle =\int_A |\widehat\psi_{\text{\rm out}}(k)|^2\, dk\ . $$ Thus in this case h.4 holds true for all $\psi_0\in{\Cal H}_c$.  Therefore hypotheses h.3 and h.4 can be interpreted as a weaker  substitute for existence and completness of wave operators. \v For explicit conditions on $\psi_0$ and $V$ ensuring admissibility the reader is referred to section 5.\v The above definitions permit us to extend (with the same proof) Carlen's results  (see [C4], the free case $V=0$ was already studied in [S]) to the case where the hypothesis of existence of the wave operators  is replaced by the weaker h.3:\v {\bf Theorem 3.} {\it Let $(\psi_0,V)$ be weakly admissible and let $(\Omega,\F,\F_t,X_t,\P)$ be as in Theorem 1. Then  $$ \l\, \frac{1}{t}\,X_t=p_+\qquad  \P\text{\rm -a.s.}\ ,\eqno(2.1) $$ for some random variable $p_+$. Moreover $p_+$ is $\P$-square integrable and it has,  under $\P$, the same distribution as does the quantum mechanical final  momentum $P_+$, i.e. for every Borel set $A$ one has  $$ \E\left(\,\chi_A( p_+)\,\right)=\langle\psi_0,\chi_A(P_+)\psi_0\rangle\ , $$ where $\E$ denotes expectation with respect to $\P$. } \v {\bf Proof.} The existence of the limit  $\l \frac{1}{t}\,X_t$ is proven in [C4, lemma 1]. For the convenience of the reader we reproduce here the main steps of such a proof. Defining the stochastic process $\pi_t:= \frac{1}{t}\,X_t$, one has the following stochastic differential equation: $$ d\pi_t=\frac{1}{t}\, (b(t,X_t)-\pi_t)\,dt+\frac{1}{t}\,dB_t\ . $$ This implies $$ \eqalign{ &\P\left(\,\sup_{t>T}\,\|\pi_t-\pi_T\|>\epsilon\,\right)\cr \le&\P\left(\,\int_T^{+\infty}\frac{dt}{t}\,\|(b(t,X_t)-\pi_t)\|\,>\epsilon\, \right) +\P\left(\,\sup_{t>T}\,\left\|\int_T^t\frac{1}{t}\,dB_t\right\|>\epsilon\,\right)\ . } $$ By Doob's martingale maximal inequality and Chebychev inequality the second term on the right can be estimated by $2\epsilon^{-2}T^{-1}$. As regards the first term, by the definition of $b$ one has $$ \eqalign{ &\P\left(\,\int_T^{+\infty}\frac{1}{t}\,\|(b(t,X_t)-\pi_t)\|\,dt>\epsilon\,\right)\cr \le&\frac{1}{\epsilon}\, \int_T^{+\infty}\frac{1}{t}\,\E(\,\|(b(t,X_t)-\pi_t)\|\,)\,dt\cr \le&\frac{1}{\epsilon}\, \int_T^{+\infty}\frac{1}{t}\,\E\left(\,\|(b(t,X_t)-\pi_t)\|^{2}\,\right)^{1/2}\,dt\cr \le&\frac{\sqrt 2}{\epsilon}\,\int_{T}^{+\infty} \left\|\left(-i\nabla-\frac{x}{t}\,\right)\psi_t\right\|_{L^2}\, \frac{dt}{t}\ . } $$ The above estimates and h.2 say that we can find a $T_n$ large enough that $$ \P\left(\bigcup_{s,t>T_n}\, \left\{\|\pi_t-\pi_s\|>\frac{1}{n}\right\}\right)<\frac{1}{2^n}\ . $$  Then, by Borel-Cantelli lemma, one has $$ \P\left(\bigcap_{m=1}^\infty\bigcap_{n>m}\bigcup_{s,t>T_n}\, \left\{\|\pi_t-\pi_s\|>\frac{1}{n}\right\}\right)=0\ , $$ which exactly means that $\l\pi_t$ exists $\P$-a.s.\par By a density argument $p_+$ has the same distribution as does the quantum mechanical final  momentum $P_+$ if $\E(g(p_+))=\langle\psi_0,g(P_+)\psi_0\rangle$ for all $g\in C_c^\infty(\RE^3)$. By h.3 there follows $$\E(g(p_+))= \l\E(g(\pi_t)) =\l\langle\psi_t,g(Q/t)\psi_t\rangle =\langle\psi_0,g(P_+)\psi_0\rangle\ , $$ and the proof is done.\qed \v {\bf Remark.} The proof of the above theorem shows that $$ \text{\rm h.2}\qquad\Longrightarrow\qquad \frac{1}{t}\,X_t\to p_+ \quad\text{\rm almost surely}\,, $$ $$ \text{\rm h.3}\qquad\iff\qquad \frac{1}{t}\,X_t\to p_+ \quad\text{\rm in distribution}\,. $$ {\bf Remark.} Under the stronger hypothesis \smallskip\p h.2.1) $$ \int_{t_0}^{+\infty} \left\|\left(P-\frac{Q}{t}\right)\psi_t\right\|^2_{L^2}\, dt <+\infty $$\p it is possible to prove (see [C5]) that the random variable $p_+$ generates the tail $\sigma$-algebra  $$ {\Cal T}:=\bigcap_{t>t_0}\sigma(X_s,\ s\ge t)\ . $$  This is the probabilistic analogue of the fact that in Quantum Mechanics the only scattering observables are functions of the final momentum $P_+$. However we will not need such a nice result here. \par Under hypothesis h.2.1, according to [C5], the proof of Theorem 3 becomes simpler:\par Let $\widetilde \P$ be the weak solution of the  simple stochastic differential equation $$ dX_t={1\over t}\,X_t\,dt\ +\ d\widetilde B_t\ . $$ Therefore  $$ d\left(\frac{1}{t}\,X_t\right)=-\frac{1}{t^2}\,X_t\,dt+\frac{1}{t}\,dX_t=\frac{1}{t}\, \widetilde B_t $$ and so $$ \frac{1}{t}\,X_t=\frac{1}{t_0}\,X_{t_0}+\int_{t_0}^t\frac{1}{s}\,d\widetilde B_s\,. $$ Since $$ \widetilde\E\left(\int_{t_0}^{+\infty}\frac{1}{s}\,d\widetilde B_s\right)^2= \int_{t_0}^{+\infty}\frac{ds}{s^2}<+\infty\,, $$ by Doob's martingale convergence theorem one gets $\widetilde\P\text{\rm -a.s.}$ convergence of $\frac{1}{t}\,X_t$. Thus the proof of Theorem 3 is then concluded by observing that h.2.1 implies $$ \E\left(\int_{t_0}^{+\infty}dt\,\|b(t,X_t)-X_t/t\|^2\right)<+\infty $$ and so, by [E, prop. 2.11], $\P$ is absolutely continuous with respect to $\widetilde \P$.  \vskip 20pt\p 
{\bf 3. The pathwise scattering-into-cones and flux-across-surfaces theorems.} \v From now on by an open cone $C$ we will mean a set of the kind $$\left\{\lambda x\in\RE^3\ :\ x\in\Sigma,\ \lambda> 0\right\}\,,$$ where $\Sigma$ is an open subset of the unit sphere with  $\partial\Sigma$ a finite union of $C^1$ manifolds.\v In the framework of Stochastic Mechanics, thanks to Theorem 3, the pathwise version of Dollard's scattering-into-cones  theorem (see [D]) is obvious: \v {\bf Theorem 4.} {\it Let $(\psi_0,V)$ be admissible and let $(\Omega,\F,\F_t,X_t,\P)$ be as in Theorem 1. Then for every open cone  $C$and for every ball $B_R$ of radius  $R$ one has $$ \l \chi_{C\cap B_R^c}( X_t)=\l \chi_{C}( X_t)=\chi_C( p_+) \qquad  \P\text{-a.s.}\ . $$} \smallskip\p {\bf Proof.} By Theorem 3 and h.4 $p_+\notin \partial C$, $\P$-a.s.. Thus by (2.1) $X_t$ is $\P$-a.s. definitively either in $C$ or in $\bar C^c$ for every open cone $C$.  Moreover, by (2.1) again, being $p_+\not=0$ $\P$-a.s. by h.4, we have  $$ \l \|X_t\|=+\infty\qquad  \P\text{\rm -a.s.}\ .\eqno(3.1) $$ Therefore  $$ \l \chi_{C\cap B_R^c}( X_t)= \l \chi_{C}( X_t)= \l \chi_{C}\left( \frac{1}{t}\,X_t\right)=\chi_C( p_+) \qquad  \P\text{\rm -a.s.}\ . $$\qed \v Let us now come to the flux-across-surfaces theorem.\v We would like to define the function  $$ N_{C\cap S_R}(\gamma):= N^+_{C\cap S_R}(\gamma)-N^-_{C\cap S_R}(\gamma)\ , $$ where $N^+_{C\cap S_R}(\gamma)$ (resp. $N^-_{C\cap S_R}(\gamma) $) denotes the number of outward (resp. inward) crossing by $[t_0,+\infty)\ni t\mapsto\gamma(t)$ of $C\cap S_R$, the intersection of  the cone $C$ with $S_R$, the sphere of radius $R$. The problem is that the above definition makes no sense since $\P$-a.s. the set $\{t\, :\, X_t\in C\cap S_R\}$ has no isolated point and  is uncountable. Therefore we are forced to proceed in an alternative way:\par Let us observe that if $\#\left\{t\,:\,\gamma(t)\in C\cap S_R\right\}<+\infty$ then $N_{C\cap S_R}(\gamma)$ is the total mass of the random distribution  $$ \sum_{t\in\left\{s\,:\,\gamma(s)\in C\cap S_R\right\}}c(t)\,\delta_{t}\ , $$ where $c(t)=+1$ if $t$ corresponds to an outward crossing and $c(t)=-1$ if $t$ corresponds to an inward crossing. Since $t\mapsto\gamma(t)$ is definitively either in $C$ or in $\bar C^c$ by h.4 and (2.1), if $R$ is sufficiently large  (then we will consider the limit $R\uparrow \infty$) one has  $$ \eqalign{ \sum_{t\in\left\{s\,:\,\gamma(s)\in C\cap S_R\right\}}c(t)\,\delta_{t}&= \sum_{t\in\left\{s\,:\,\gamma(s)\in (C\cap S_R)\cup(\partial C\cap B_R^c)\right\}}c(t)\,\delta_{t}\cr &=\frac{d}{dt}\, \chi_{C\cap \bar B_R^c}(\gamma(t))\,, } $$ where the derivative has to be intended in distributional sense.  The advantage of this rewriting is that for every path $\gamma$ the distribution  $\frac{d}{dt}\, \chi_{C\cap \bar B_R^c}(\gamma(t))$ is well defined. \v {\bf Definition.} Given an open domain $D$, we define the random distribution  $$\mu_D:\Omega\to {\Cal D}'(\RE)$$  by $$ \mu_D(\gamma):={d\over{dt}}\,\chi_D(\tilde\gamma(t))\, ,\qquad \tilde\gamma(t):=\cases \gamma(t),&\text{for $t\ge t_0$}\\ \gamma(t_0),&\text{for $t<t_0$}\, ,\endcases $$ i.e. for every test function $\phi\in {\Cal D}(\RE)\equiv C_c^\infty(\RE)$ $$ \langle\mu_D(\gamma),\phi\rangle :=-\chi_{D}(\gamma(t_0))\,\phi(t_0)-\int_{t_0}^{+\infty}dt\, \chi_D(\gamma(t))\,\dot\phi(t)\ . $$ Note that supp$[\mu_D(\gamma)]=\gamma^{-1}(\partial D)$.  In the case $\mu_D(\gamma)\in {\Cal E}'(\RE)$, i.e. it has compact support, we define as usual its mass by $$ M_{ D}(\gamma):=\langle\mu_D(\gamma),\phi_\gamma\rangle\ ,$$ where $\phi_\gamma$ is a test function such that $\phi_\gamma=1$ on a neighbourhood of supp[$\mu_D(\gamma)$]. \v By the previous definition and by Theorem 3 we have then the following  pathwise version of the flux-across-surfaces theorem:\v  {\bf Theorem 5.} {\it Let $(\psi_0,V)$ be admissible and let $(\Omega,\F,\F_t,X_t,\P)$ be as in Theorem 1. Then  $$\mu_{C\cap \bar B^c_R}\in {\Cal E}'(\RE)\qquad  \P\text{-a.s.}\,.$$ and, defining $N_{C\cap S_R}:=M_{C\cap \bar B^c_R}$, one has  $$ \lim_{R\uparrow\infty}\,N_{C\cap S_R}=\chi_C(p_+)\qquad  \P\text{-a.s.}\, . $$ } \smallskip\p {\bf Proof.} Let $$ \tau_R(\gamma):=\sup\left\{t\in\RE\ :\ X_t(\gamma)\in\partial(C\cap\bar B^c_R)\right\}\ . $$ By (3.1) and since $X_t$ is $\P$-a.s. definitively either in $C$ or in $\bar C^c$ by h.4 and Theorem 3, one has  $\tau_R<+\infty$, $\P$-a.s.. Thus $\mu_{C\cap \bar B_R^c}\in {\Cal E}'(\RE)$, $\P\text{-a.s.}$, being  supp$[\mu_{C\cap \bar B^c_R}]\subseteq[t_0,\tau_R(\gamma)]$.\par  Let $\phi_\gamma\in {\Cal D}(\RE)$ such that $\phi_\gamma=1$ on a neighbourhood of $[t_0,\tau_R(\gamma)]$.  By the definition of $\mu_{C\cap \bar B^c_R}$ one has  $$\eqalign{ \langle\mu_{C\cap \bar B^c_R}(\gamma),\phi_\gamma\rangle =&-\chi_{{C\cap \bar B^c_R}}(\gamma(t_0))-\chi_C(p_+(\gamma)) \int_{\tau_R(\gamma)}^{+\infty}dt\,\dot\phi_\gamma(t)\cr =&-\chi_{{C\cap \bar B^c_R}}(\gamma(t_0))+\chi_C(p_+(\gamma))\ , } $$ and the thesis then immediately follows by taking the limit $R\uparrow\infty$. \qed \vskip 20pt\p {\bf 4. The scattering-into-cones and flux-across-surfaces theorems in Quantum Mechanics.} \v By taking expectations in Theorem 4 and by dominated convergence theorem one immediately obtains Dollard's theorem:  \v {\bf Theorem 6.} {\it For every open cone $C$, every ball $B_R$ of radius  $R$, and for every admissible couple  $(\psi_0,V)$, one has  $$ \l \int_{C\cap B_R^c}dx\,|\psi_t(x)|^2 =\l \int_{C}dx\,|\psi_t(x)|^2 =\langle\psi_0,\chi_C(P_+)\psi_0\rangle\ . $$  } In order to prove the flux-across-surfaces theorem we need now to compute the expectation of $\mu_{C\cap\bar B^c_R}$. To this end we state the following\v {\bf Theorem 7.} {\it Let $\psi_t$  and $\P$ be as in Theorem 1, with $\psi_0\in H^2(\RE^3)$ and $V$ a $(-\Delta)$-operator-bounded  potential, with  relative bound smaller than one. For every open domain $D$, with $\partial D$ a finite union of $C^1$ manifolds, and for  every test function $\phi$ one has $$ \E(\langle\mu_{D},\phi\rangle)=-\int _{t_0}^{+\infty}dt\,\phi(t)\int_{\partial D}d\sigma(x)\,J^{\psi_t}(x) \cdot n(x) \ , $$ where $n$ denotes the outward unit normal vector along $\partial D$  and $\sigma$ is the surface measure.} \p\smallskip\p{\bf Proof.} Since $|\psi_t|^2$ is the density of $X_t$ under $\P$,   $$ \E\langle\mu_D,\phi\rangle=-\phi(t_0)\int_Ddx\,|\psi_{t_0}(x)|^2 -\int_{t_0}^{+\infty}dt\,\dot\phi(t)\int_Ddx\,|\psi(t,x)|^2\ . $$ Since $\psi_t$ solves the Schr\"odinger equation, one has (see e.g. [C1]), for all $f\in C_b^1(\RE^3)$ and for a.e. $t$, the continuity equation $$ \frac{d}{dt}\int_{\RE^3}dx\,  |\psi_t(x)|^2\,f(x)=\int_{\RE^3}dx\,J^{\psi_t}(x)\cdot\nabla f(x)\ . $$ Since $\psi_t\in H^2(\RE^3)$ by our hypotheses on $\psi_0$ and $V$, $\nabla J^{\psi_t}$ is an  integrable function. Therefore one has, integrating by parts,  for all $f\in C_b^1(\RE^3)$  $$ \eqalign{ &\int_{t_0}^{+\infty}dt\,\dot\phi(t)\int_{\RE^3}dx\,  |\psi_t(x)|^2\,f(x)\, =-\phi(t_0)\int_{\RE^3}dx\,|\psi_{t_0}(x)|^2f(x)\cr &+\int_{t_0}^{+\infty}dt\,\phi(t) \int_{\RE^3}dx\,\nabla J^{\psi_t}(x) f(x)\ . } $$ Taking now a uniformly bounded sequence $\left\{f_n\right\}_1^\infty\subset C_b^1(\RE^3)$, pointwise converging to $\chi_D$, by the dominated convergence theorem one obtains $$\eqalign{ &\int_{t_0}^{+\infty}dt\,\dot\phi(t)\int_{D}dx\,  |\psi_t(x)|^2\cr =&-\phi(t_0)\int_Ddx\,|\psi_{t_0}(x)|^2+\int_{t_0}^{+\infty}dt\, \phi(t)\int_{D}dx\,\nabla J^{\psi_t}(x)\ . } $$ Since $\psi_t\in H^2(\RE^3)$, one has $\nabla\psi_t\in H^1(\RE^3)$, so that both $\psi_t$ and $\nabla\psi_t$ have traces in $L^2(\partial D)$ (see e.g. [B, chap. 5]).  Thus $J^{\psi_t}$ has a trace in $L^1(\partial D)$ by  $$\|J^{\psi_t}\|_{L^1(\partial D)}\le  \|\psi_t\|_{L^2(\partial D)}\|\nabla\psi_t\|_{L^2(\partial D)}\le  c\,  \|\psi_t\|_{H^1(D)}\|\nabla\psi_t\|_{H^1(D)}\,, $$  and the proof is then concluded by the Gauss-Green theorem.\qed \v {\bf Definition 8.} The admissible couple  $(\psi_0,V)$ is said to be {\it strongly admissible} if \smallskip\p  h.5) $\psi_0\in H^2(\RE^3)$, $V$ is a $(-\Delta)$-operator-bounded  potential and  $$ \int_{t_0}^{\infty}dt\,\left\|\theta(Q)\psi_t\right\|_{H^1}\, \left\|\left(P-\frac{Q}{t}\right)\psi_t\right\|_{H^1}<+\infty $$ where $\theta\in C^2_b(\RE^3;\C)$ such that $\theta=1$ on a neighbourhood of $\partial C\cap B_R^c$ for some $R>0$. \v  In the next section we will give explicit conditions on $\psi_0$ and $V$ ensuring strong admissibility. \v By combining Theorems 5 and 7 the flux-across-surfaces theorem now follows:\v {\bf Theorem 9.} {\it For every open cone $C$ and for every strongly admissible couple  $(\psi_0,V)$ one has  $$ \lim_{R\uparrow \infty}\,\lim_{T\uparrow \infty}\, \int_{t_0}^{T}dt\int_{{C\cap S_R}} d\sigma(x)\,J^{\psi_t}(x)\cdot n(x) =\langle\psi_0,\chi_C(P_+)\psi_0\rangle\,. $$ } \p\smallskip\p{\bf Proof.} By pointwise approximating,  on the compact interval $[t_0,\tau_R(\gamma)]$, $t\mapsto \gamma(t)$ with a sequence of polynomials paths,  the wildly oscillating function $t\mapsto \chi_{C\cap \bar B^c_R}(\gamma(t))$ can be pointwise approximated with a sequence  $\left\{\chi_n\right\}_1^\infty$ of characteristic functions of finite unions $\displaystyle \bigcup_{k=0}^{m(n)} [s_k^{(n)},t^{(n)}_{k}]$ of disjoint intervals. Therefore one obtains $$\eqalign{ |\langle \mu_{C\cap \bar B^c_R}(\gamma),\phi\rangle|\le & \chi_{C\cap\bar B^c_R}(\gamma(t_0))\,|\phi(t_0)|+\lim_{n\uparrow\infty} \sum_{k=0}^{m(n)}\left|\int_{s_k^{(n)}}^{t^{(n)}_{k}}dt\,\dot\phi(t) \right|\cr =&\chi_{C\cap\bar B^c_R}(\gamma(t_0))\,|\phi(t_0)|+\lim_{n\uparrow\infty} \sum_{k=0}^{m(n)}\left|\phi(t^{(n)}_{k}) -\phi(s_k^{(n)})\right|\cr \le &\chi_{C\cap\bar B^c_R}(\gamma(t_0))\,|\phi(t_0)|+{\text{ \rm var}}(\phi) \ . } $$ Now let us note that in Theorem 5 we can alternatively define $N_{C\cap S_R}$ by  $$ N_{C\cap S_R}(\gamma):= \lim_{n\uparrow\infty}\,\lim_{m\uparrow\infty}\, \langle\mu_{C\cap \bar B^c_R}(\gamma),\phi_{n,m}\rangle $$  where $\{\phi_{n,m}\}_{n,m\ge 1}$ is a double  sequence of test functions such that $\phi_{n,m}=1$ on $[t_0,n]$, $\phi_{n,m}\to\chi_{[t_0,n]}$ pointwise.  Then if we choose such test functions $\phi_{n,m}$ in such a way that their variation is bounded uniformly in $n$ and $m$,  by the dominated convergence theorem and by Theorem 5 one has   $$\eqalign{ &\langle\psi_0,\chi_C(P_+)\psi_0\rangle= \E(\chi_C(p_+))\cr =& \lim_{R\uparrow\infty}\,\lim_{n\uparrow\infty}\,\lim_{m\uparrow\infty}\, \E(\langle\mu_{C\cap B_R^c},\phi_{n,m}\rangle)\cr =& \lim_{R\uparrow \infty}\,\lim_{n\uparrow\infty}\, \int_{t_0}^{n}dt\int_{(C\cap S_R)\cup ( \partial C\cap B_R^c)} d\sigma(x)\,J^{\psi_t}(x) \cdot n(x)\ . } $$ The proof is then conlcuded by proving that   $$ \lim_{R\uparrow \infty}\,\lim_{n\uparrow\infty}\, \int_{t_0}^{n}dt\int_{ \partial C\cap B_R^c} d\sigma(x)\,J^{\psi_t}(x) \cdot n(x)=0\,.\eqno(4.1) $$ Since $n\cdot x=0$ on $\partial C$ and $\|J^{\psi_t}\|\le \|\psi^*_t\nabla\psi_t\|=\|\psi_tP\psi_t\|$, one has   $$ \eqalign{ &\left|\int_{t_1}^{n}dt\int_{ \partial C\cap B_R^c} d\sigma(x)\, J^{\psi_t}(x) \cdot n(x)\right|\cr \le &\int_{t_1}^{n}dt\int_{ \partial C\cap B_R^c} d\sigma(x)\, \|\psi_t(x)P\psi_t(x)\|\cr =&\int_{t_1}^{n}dt\int_{ \partial C\cap B_R^c} d\sigma(x)\, \left\|\psi_t(x)\left(P-\frac{Q}{t}\right)\psi_t(x)\right\|\,.} $$ Thus, since $\psi_tP\psi_t\in L^1(\partial C)$, by the monotone convergence theorem, (4.1) follows from  $$ \int_{t_0}^{\infty}dt\int_{ \partial C\cap B_R^c} d\sigma(x)\, \left\|\psi_t(x)\left(P-\frac{Q}{t}\right)\psi_t(x)\right\|<+\infty\eqno(4.2) $$ for some $R>0$. By trace estimates on functions in $H^1(\RE^3)$ of the kind  $$\|\cdot\|_{L^2(\partial C)}\le c\, \|\cdot\|_{H^1(\RE^3)}\,,$$ (see e.g. [B, chap. 5]) one has $$ \eqalign{ &\int_{ \partial C\cap B_R^c} d\sigma(x)\, \left\|\psi_t(x)\left(P-\frac{Q}{t}\right)\psi_t(x)\right\|\, \cr &\left(\int_{ \partial C\cap B_R^c} d\sigma(x)\, |\theta(x)\psi_t(x)|^2\right)^{1/2} \left(\int_{ \partial C\cap B_R^c} d\sigma(x)\, \left\|\left(P-\frac{Q}{t}\right)\psi_t(x)\right\|^2\right)^{1/2}\cr &\le c\,\left\|\theta(Q)\psi_t\right\|_{H^1}\, \left\|\left(P-\frac{Q}{t}\right)\psi_t\right\|_{H^1}\,, } $$ so that (4.2) is a consequence of h.5. \qed    \vskip 20pt\p {\bf 5. On the admissibility conditions.} \v When $V=0$, $\psi_0\in H^2(\RE^3)$ and $|Q|\psi_0\in L^2(\RE^3)$, by the explicit espression for $e^{it\Delta}\psi_0$ one has  $$ \left\|\left(P-\frac{1}{t}\,Q\right)\psi_t\right\|_{L^2} =\frac{1}{t}\,\|Q\psi_0\|_{L^2}\ , $$ and therefore the free case satisfies the admissibility hypothesis h.2-h.4 (with $P_+=P$).\par Now we come to the interacting case. The condition h.3 gives no trouble: it follows from fairly general hypotheses on the potential function $V$. Indeed by [DG, thm. 4.4.1], $$ V(-\Delta+\uno)^{-1}\quad{\text{is compact}}\eqno (5.1) $$ and $$ \int_1^{+\infty}dR\, \left\|(-\Delta+\uno)^{-1}\nabla V\,\chi_{[1,+\infty)}\left(\frac{\|x\|}{R}\right) \,(-\Delta+\uno)^{-1}\right\|_{L^2,L^2}<+\infty\, ,\eqno(5.2) $$ imply (a stronger version of) h.3. By [HS, thm. 14.9], if for all $\epsilon>0$ we can decompose $V=V_1+V_2$ with $V_1\in\L$ and $V_2\in L^\infty(\RE^3)$, with $\|V_2\|_\infty<\epsilon$, then (5.1) holds true. If, outside some ball, $V$ is differentiable with its first derivatives decaying at infinity faster than $\|x\|^{-1}$ then condition (5.2) follows.\par  As regards the condition h.2, by the proof of [C4, lemma 4] one has  $$ \left\|\left(P-\frac{Q}{t}\right)\psi_t\right\|_{L^2} \le \frac{1}{t}\,\|(P-Q)\psi_{1}\|_{L^2} +\frac{1}{t}\,\int_{1}^t ds\,s\,\|\psi_s\nabla V\|_{L^2}\ . $$ By (5.1) $V$ is infinitesimally $(-\Delta)$-operator-bounded (see e.g. [HS, thm. 14.2]) and so (see [C1, thm. 2.1(iv)]) $\|(-i\nabla-x)\psi_{t}\|_{L^2}<+\infty$ for all $t$ if $$\psi_0\in H^2(\RE^3)\, ,\qquad |Q|\psi_0\in L^2(\RE^3)\ .$$  Therefore h.2 follows from  $$ \|\psi_t\nabla V\|_{L^2}\le c\,(1+|t|\,)^{-\sigma}\, ,\qquad \sigma>1\,. \eqno(5.3) $$  \v  We introduce the notations $\langle x\rangle$ for the function $(1+\|x\|^2)^{1/2}$ and $\langle Q\rangle$ for the corresponding multiplication operator.\v  In the case $\langle Q\rangle ^s\nabla V\in L^\infty(\RE^3)$ and $\langle Q\rangle^s\psi_0\in L^2(\RE^3)$ for some $s$, (5.3) then follows from  $$ \left\|\langle Q\rangle^{-s}e^{-itH}\langle Q\rangle^{-s}\right\|_{L^2,L^2}\le c\,(1+|t|\,)^{-\sigma}\,.\eqno(5.4) $$ Such a kind of estimates were obtained in many paper about propagation estimates for solution of Schr\"odinger equations (see e.g. [ACS], [CP], [JK], [JMP], [JSS]). For example, by [CP, thm.1], one obtains that h.2 holds true under the following hypotheses: $$ \psi_0\in H^2(\RE^3)\,,\qquad\langle Q\rangle^s\psi_0\in L^2(\RE^3)\,,\qquad \phi(H)\psi_0=\psi_0\,,\eqno(5.5) $$ $$ V=V_S+V_L\,,\qquad V_S\in C^1(\RE^3)\,,\quad V_L\in C^{k+3}(\RE^3)\,,\eqno(5.6) $$ $$ \|D^\alpha V_S(x)\|\le c\,\langle x\rangle^{-2k-|\alpha|-\epsilon} \, ,\qquad|\alpha|\le 1\,,\eqno(5.7) $$ $$  \|D^\alpha V_L(x)\|\le  c\,\langle x\rangle^{-|\alpha|-\epsilon}\, ,\qquad |\alpha|\le k+1\, ,\eqno(5.8) $$ where $\phi\in C^{\infty}(0,+\infty)$ is equal to zero on a (arbitrarily small) neighbourhood of zero and   $$\epsilon>0\,,\qquad k\ge 3\,, \qquad 1<s\le k\,,\qquad  s\left(1-\frac{1}{k}\right)>1\eqno(5.9)$$ ($s(1-{1}/{k})>3/2$ gives h.2.1).\v Note that under these conditions (5.1) and (5.2) hold true.  Moreover, by [HS, thm. 16.1]  there are no strictly positive eigenvalues and, by [DG, thm. 4.7.1], one has also existence and completeness of the (modified) wave operators, so that, for every Borel set $A$, $$ \langle\psi_0,\chi_A(P_+)\psi_0\rangle= \int_A dk\,|\widehat\psi_{\text{\rm out}}(k)|^2\, . $$ Thus h.4 holds true and in conclusion  $$ \text{\rm (5.5)-(5.9)} \quad\Longrightarrow\quad \text{\it admissibility}\,.$$ \v {\bf Remark.} By [JK, thm. 10.3] the low energy cutoff hypothesis can be removed when $0$ is neither an eigenvalue nor a resonance, $\epsilon>3$ and $s>5/2$.  If $0$ is not an eigenvalue but is a resonance then $\psi_0$ has to be orthogonal to the function corresponding to the resonance, otherwise in (5.4) one has  $\sigma=1/2$ (see [JK, thm. 10.5]). \v As regards strong admissibility, i.e  hypothesis h.5,  the main point in the paper [AP] by Amrein and Pearson was just to find the conditions on $\psi_0$ and $V$ leading to such an hypothesis. By using again (5.4) and commutator estimates, by [AP, lemmata 5-8] one obtains  $$ \text{\rm (5.5)-(5.9) with $s>5/3$ and $\epsilon>2/3$} \quad\Longrightarrow\quad \text{\it strong admissibility}\,.$$  In [AP, section 6] it is then shown how to avoid regularity hypotheses on  the short range component of $V$. However in this situation  the hypotheses on the initial state $\psi_0$ become less transparent. Indeed  there one requires $W_+^*\psi_0= \varphi(H_1)W_+^*\psi_0$, $\langle Q\rangle^s W_+^*\psi_0\in L^2(\RE^3)$, $s>2$, $\varphi\in C_c(0,+\infty)$,  $H_1:=-\Delta+V_1$, $V_1$ the smooth part of $V$, $W_+$ the relative wave operator $W_+:=\lim_{t\uparrow\infty}e^{-itH} e^{itH_1}$.    \v We conclude the section by listing the conditions on the couple $(\psi_0,V)$ used in other papers (beside the already quoted [AP])  in order to obtain the flux-across-surfaces theorem (${\Cal S}(\RE^3)$ denoting the space of functions of rapid decrease):  \v 1. In [DDGZ1] is it assumed that $V=0$ and $\psi_0\in{\Cal S}(\RE^3)$. \smallskip\p 2. In [AZ] it is assumed that $\langle Q\rangle^s\psi_{\text{\rm out}}\in L^2(\RE^3)$, $s>5/2$,  $\psi_{\text{\rm out}}=\varphi(-\Delta)\psi_{\text{\rm out}}$, $\varphi\in C_c^{\infty}(0,+\infty)$, $V$ either has local singularities and decays faster than $\|x\|^{-2}$ at infinity or is in $C^4(\RE^3)$ and decays faster than $\|x\|^{-1}$ (in this case  $\widehat\psi_{\text{\rm out}}$ has to be in  $C^4_c(\RE^3\backslash\left\{0\right\})$).  By [JN], when $V$ is smooth, the condition on the outgoing state  $\psi_{\text{\rm out}}$ is implied by a similar one  (with $s>7/2$) on $\psi_0$.  \smallskip\p 3. In [TDM-B] it is assumed that $\psi_{\text{\rm out}}\in{\Cal S}(\RE^3)$,  that $V\in\L$ is locally H\"older continuous except at a finite number of points and is decaying faster than $\|x\|^{-4}$, and that $0$ is neither an eigenvalue nor a resonance. No energy cutoff condition on $\psi_0$ is required. \smallskip\p 4. In [DPa] the  results in [TDM-B] are extended to the case in which $0$ is either a zero-energy eigenvalue or resonance. There it is assumed that  $\psi_0\in {\Cal S}(\RE^3)$, $V\in\L$ is locally H\"older continuous except at a finite number of points and is decaying faster than $\|x\|^{-n}$ for  all $n\in{\Bbb N}$, $\widehat\psi_{\text{\rm out}}\in C^5(\RE^3\backslash\left\{0\right\})$ and  $\|D^\alpha \widehat\psi_{\text{\rm out}}(k)\|\le c\,\langle k\rangle ^{-3-|\alpha|-\epsilon}$, $|\alpha|\le 5$, $\epsilon>0$, $\|k\|\ge K_\alpha>0$. \smallskip\p 6. In [PT] and [DPa] the flux-across-surfaces theorem is proven in the case in which $\psi_0\in {\Cal S}(\RE^3)$ and $H$ is the self-adjoint operator describing the Laplacean with a delta point interaction.  \vskip 40pt\p {\bf Acknowledgments.} We thank Sergio Albeverio, Eric Carlen, Gianfausto Dell'Antonio for stimulating discussions and Detlef D\"urr for a critical reading of an early draft of the manuscript.    
 \vskip 40pt \centerline{\bf References} \vskip 10pt\p \item{[ACS]} Amrein, W.O., Cibils, M.B., Sinha, K.B.: Configuration  Space Properties of the S-Matrix and Time Delay in Potential Scattering. {\it Ann. Inst. Henri Poincar\'e} {\bf 47}, 367-382 (1987) \item{[AP]} Amrein, W.O., Pearson, D.B.: Flux and Scattering into Cones for Long Range and Singular Potentials. {\it Journal of Physics A}, {\bf 30}, 5361-5379 (1997) \item{[AZ]} Amrein, W.O., Zuleta, J.L.: Flux and Scattering into Cones in Potential Scattering. {\it Helv. Phys. Acta} {\bf 70}, 1-15 (1997) \item{[BDGPZ]} Berndl, K., D\"urr, D., Goldstein, S., G. Peruzzi, Zangh\`\i, N.: On the Global Existence of Bohmian Mechanics.  {\it Commun. Math. Phys.} {\bf 173}, 647- (1995) \item{[B]} Burenkov, V.I.: {\it Sobolev Spaces on Domains.} Stuttgart, Leipzig: Teubner 1998 \item{[C1]} Carlen, E.: Conservative Diffusions. {\it Commun. Math.  Phys.} {\bf 94},  293-315 (1984) \item{[C2]} Carlen, E.: Existence and Sample Path Properties of the Diffusions  in Nelson's Stochastic Mechanics. In: Albeverio, S. et al. (eds.)  {\it Stochastic Processes - Mathematics and Physics.  Lecture notes in Mathematics, Vol. 1158}, pp. 25-51. Berlin, Heidelberg, New  York: Springer 1985 \item{[C3]} Carlen, E.: Progress and Problems in Stochastic Mechanics.  In: Gielerak, R., Karwowski, W. (eds.) {\it Stochastic Methods in Mathematical  Physics}, pp. 3-31.  Singapore: World Scientific 1989 \item{[C4]} Carlen, E.:  Potential Scattering in Stochastic Mechanics.  {\it Ann. Inst. Henri Poincar\'e} {\bf 42}, 407-428  (1985). \item{[C5]} Carlen, E.: The Pathwise Description of Quantum Scattering  in Stochastic Mechanics. In: Albeverio, S. et al. (eds.)  {\it Stochastic Processes in Classical and  Quantum Systems. Lecture Notes in Physics, Vol.  262}, pp. 139-147. Berlin, Heidelberg, New York: Springer 1986.  \item{[CP]} Cycon, H.L., Perry, P.A.: Local Time-Decay of High Energy Scattering States for The Schr\"odinger Equation. {\it Math.Z.} {\bf 188}, 125-142 (1984) \item{[CNS]} Combes, J.M., Newton, R.G., Shtokhamer, R.: Scattering into Cones and Flux Across Surfaces. {\it Phys. Rev. D}  {\bf 11}, 366-372 (1975) \item{[DDGZ1]} Daumer, M., D\"urr, D., Goldstein, S., Zangh\`\i, N.: On the Flux-Across-Surfaces Theorem. {\it Lett. Math. Phys.} {\bf 38}, 103-116 (1996) \item{[DDGZ2]} Daumer, M., D\"urr, D., Goldstein, S., Zangh\`\i, N.: On the Quantum Probability Flux through Surfaces. {\it J. Stat. Phys.} {\bf 88}, 967-977 (1997) \item{[DPa]} Dell'Antonio, G.F., Panati, G.: Zero-Energy Resonances and he Flux-Across-Surfaces Theorem. Preprint (2001) \item{[DP]} Dell'Antonio, G.F., Posilicano, A.: Convergence of Nelson Diffusions. {\it Comm. Math. Phys.} {\bf 141}, 559-576, 1991 \item{[DG]} Derezi\'nski, J., G\'erard, C.: {\it Scattering Theory of Classical and Quantum N-Particle Systems.} Berlin, Heidelber, New York: Springer 1997  \item{[D]} Dollard, J.D.: Scattering into Cones I, Potential Scattering. {\it Comm. Math. Phys.} {\bf 12}, 193-203 (1969) \item{[DGZ]} D\"urr, D., Goldstein, S., Zangh\`\i, N.: Quantum Equilibrium and the Origin of Absolute Uncertainity {\it J. Stat. Phys.} {\bf 67}, 843-907 (1992) \item{[E]} Ershov, M.: On the Absolute Continuity of Measures Corresponding to  Diffusion Processes. {\it Theory of Prob. and Appl.} {\bf 17}, 169-174 (1972) \item{[GM]} Guerra, F., Morato, L.M.: Quantization of Dynamical Systems ans Stochastic Control Theory. {\it Phys. Rev.} {\bf D27}, 1774-1786 (1983) \item{[HS]} Hilsop, P.D., Sigal, I.M.: {\it Introduction to Spectral Theory.} Berlin, Heidelberg, New York: Springer 1996 \item{[JK]} Jensen, A., Kato, T.: Spectral Properties of Schr\"odinger Operators and Time-Decay of the Wave Functions. {\it Duke Math. J.} {\bf 46}, 583-611 (1979) \item{[JMP]} Jensen, A., Mourre, \'E., Perry, P.: Multiple Commutator Estimates and Resolvent Smoothness in Quantum Scattering Theory.  {\it Ann. Inst. Henri Poincar\'e} {\bf 41}, 207-225 (1984) \item{[JN]} Jensen, A., Nakamura, S.: Mapping Properties of Wave and Scattering Operators for Two-Body Schr\"odinger Operators. {\it Lett. Math. Phys.} {\bf 24}, 295-305 (1992)   \item{[JSS]} Journ\'e, J.L., Soffer, A., Sogge, C.D.: Decay estimates for Schr\"odinger operators.  {\it Commun. Pure Appl. Math.} {\bf 44}, 573-604 (1991) \item{[N1]} Nelson, E.: Derivation of the Schr\"odinger Equation from  Newtonian Mechanics. {\it Phys. Rev.} {\bf 150}, 1079-1085 (1966) \item{[N2]} Nelson, E.: {\it Dynamical Theories of Brownian Motion.}  Princeton: Princeton Univ. Press 1967 \item{[N3]} Nelson, E.: {\it Quantum Fluctuations.}  Princeton: Princeton Univ. Press 1985 \item{[PT]} Panati, G., Teta, A.: The Flux-Across-Surfaces for a Point  Interaction Hamiltonian. In: Gesztesy, F., et al. (eds.)   {\it Stochastic Processes, Physics and Geometry: New Interplays. II: A Volume in Honor of Sergio Albeverio.} Providence, Rhode Island: Am. Math. Soc. 2000 \item{[S]} Shucker, D.: Stochastic Mechanics of Systems with Zero Potential. {\it J. Funct. Anal.} {\bf 38}, 146-155 (1980) \item{[TDM-B]} Teufel, S., D\"urr, D., M\"unch-Berndl, K.: The Flux-Across-Surfaces Theorem for Short Range Potentials and Wave Functions without Energy Cutoffs. {\it J. Math. Phys.} {\bf 40}, 1901-1922 (1999)  \end